\documentclass[preprint,aps,showpacs,preprintnumbers,amsmath,amssymb]{revtex4}

\usepackage{graphicx}% Include figure files
\usepackage{dcolumn}% Align table columns on decimal point
\usepackage{bm}% bold math
\begin{document}

%%%%\preprint{APS/123-QED}

\title{
Measurement and control of spatial qubits generated by passing photons through double-slits
}

\author{Gen Taguchi}\email{gentgch@hiroshima-u.ac.jp}\author{Tatsuo Dougakiuchi}\author{Nobuaki Yoshimoto}\author{Katsuya Kasai}\author{Masataka Iinuma}\author{Holger F. Hofmann}\author{Yutaka Kadoya}
\affiliation{%
Graduate School of Advanced Sciences of Matter, Hiroshima University, 
Higashihiroshima, 739-8530, Japan
}%

\date{\today}% It is always \today, today,

\begin{abstract}
We present an experimental study of the non-classical correlations
of a pair of spatial qubits formed by passing two down-converted photons
through a pair of double slits. After confirming the entanglement 
generated in our setup by quantum tomography using separate measurements
of the slit images and the interference patterns, we show that the complete Hilbert space of the spatial qubits can be accessed by measurements
performed in a single plane between the image plane and the focal plane
of a lens. Specifically, it is possible to obtain both the which-path 
and the interference information needed for quantum tomography in
a single scan of the transversal distribution of photon coincidences.
Since this method can easily be extended to multi-dimensional systems,
it may be a valuable tool in the application of spatial qudits
to quantum information processes. 
\end{abstract}

% PACS, the Physics and Astronomy Classification Scheme.
%
%\keywords{Suggested keywords}%Use showkeys class option if keyword display desired
%
\maketitle
%
%%%%%%%%%%%%%%%%%%%%%%%%%%%%%%%%%%%%%%%%
\section{Introduction}
%%%%%%%%%%%%%%%%%%%%%%%%%%%%%%%%%%%%%%%%
\par
Quantum information science makes use of the entanglement
between well-defined two dimensional qubits or d-dimensional
qudits to perform tasks that could not be performed by a
corresponding classical system. In optical implementations
of quantum information technologies, a commonly used source 
of entanglement is the generation of photon pairs by 
spontaneous parametric down-conversion (SPDC) 
\cite{Ou1988,Kwiat1995}. 
%%%%%%%%%%%%%%%%%%%%%%%%%%%%%%%%%%%%%%%%
\par
In many applications, photon polarization is used
to define a qubit, since it provides a natural two level system that
is comparatively easy to control by using birefringent optical
elements. However, photon pairs produced by SPDC are also
entangled in their spatial and temporal degrees of freedom
\cite{Franson1989,Strekalov1995,Howell2004,D'Angelo2004}.
Since these degrees of freedom are naturally continuous, it is
necessary to introduce additional constraints in order to 
define a qubit or qudit system. The advantage of this approach is
that it is fairly easy to extend the method to higher dimensions,
a possibility that may simplify some quantum information processes
\cite{Kaszlikowski2000,Joo2007,Lanyon2008a}.
%%%%%%%%%%%%%%%%%%%%%%%%%%%%%%%%%%%%%%%%
\par
One widely used method of defining spatial qudits by
discretizing the transversal degrees of freedom is the 
selection of angular momentum eigenstates corresponding 
to photons in Gauss-Laguerre modes \cite{Mair2001,Vaziri2003,Oemrawsingh2004,Langford2004,Lanyon2008b}. However, 
it is in principle not necessary to base the selection of modes
on symmetries, since the spatial entanglement applies to 
arbitrary selections of orthogonal modes. It is therefore 
possible to define qudits by simply selecting a sufficiently
narrow spatial ``pixel'' for each basis state of the qudit 
\cite{O'Sullivan-Hale2005}. Alternatively, it is possible to
concentrate on only one spatial dimension. As was demonstrated by
Neves and co-workers, entangled spatial qudits can then be obtained 
by using the familiar slit arrays used in basic demonstrations of optical 
interference \cite{Neves2005}.
%%%%%%%%%%%%%%%%%%%%%%%%%%%%%%%%%%%%%%%%
\par
An interesting and important aspect of the use of multi-slits to
define spatial qudits is the fact that the photons continue to
propagate in continuous free space after their wavefunction
has been projected into a two dimensional Hilbert space
\cite{Lima2006}. This means that the spatial qudits continue
to exist in the infinite dimensional Hilbert space of transverse
position and momentum. Specifically, the only directly observable
physical property of the qudits is their transverse position,
so that the measurement and control of the qudit states has to
be based on the evolution of their spatial wavefunction. 
It is therefore of great interest to explore the possibilities
of preparing and characterizing different multi-slit qudit states
using their propagation in space.
%%%%%%%%%%%%%%%%%%%%%%%%%%%%%%%%%%%%%%%%
\par
In this paper, we present a thorough investigation of the entanglement
between a pair of double-slit qubits based on measurements of their
correlated spatial patterns. In particular, we show that measurements
performed between the focal and image planes of a lens simultaneously
provide which-path information that distinguishes between the slits
and interference information that identifies the quantum coherence 
between the slits. It is therefore possible to scan the whole
surface of the qubit Bloch sphere by varying the detector position
in this intermediate plane. We can use this method to prepare arbitrary
superposition states, and to perform quantum tomography of the
states thus prepared by scanning only a single transverse pattern.
%%%%%%%%%%%%%%%%%%%%%%%%%%%%%%%%%%%%%%%%
\par
The rest of the paper is organized as follows.
In section \ref{Sec2}, we present the experimental setup and
report the results of quantum tomography performed by 
measuring count rates corresponding to eigenstates of the
Pauli operators in the image and focal planes.
In section \ref{Sec3}, we analyze the effects of a measurement
between the focal and image planes on the spatial qubit and
show how it can be used for qubit preparation and for
single-scan tomography.
In section \ref{Sec4}, we present the experimental results 
of single-scan tomography for conditional states prepared
by measurements in the other arm and reconstruct the density
matrix of the entangled state once more from these results.
The results are compared with those from the Pauli operator
measurements, and possible experimental problems are discussed.
Section \ref{Sec5} concludes the paper.
%
%
%%%%%%%%%%%%%%%%%%%%%%%%%%%%%%%%%%%%%%%%
\section{Generation of entangled spatial qubits\label{Sec2}}
\begin{figure}
\includegraphics[width=0.918\linewidth,height=0.216\linewidth]{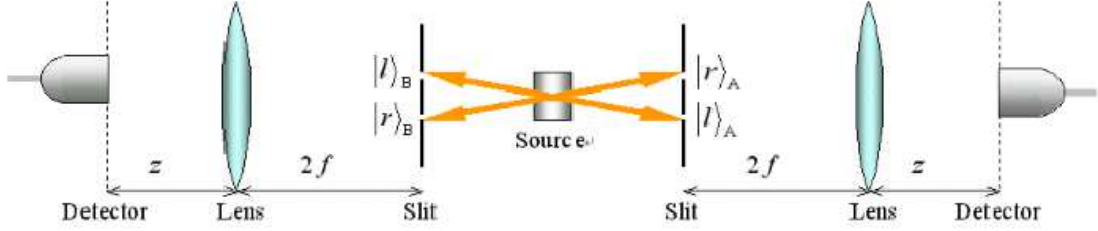}
\caption{\label{fig:Concept} Definition of the entangled qubits. 
The slits in each arm A, B define the basis states $\{|l\rangle, |r\rangle\}$ corresponding to the slit a photon passes through. 
}
\end{figure}

%%%%%%%%%%%%%%%%%%%%%%%%%%%%%%%%%%%%%%%%
In this work, we use double-slits to define the spatial qubits. 
It is convenient to express the qubit states in the $\{|l\rangle, |r\rangle\}$ basis corresponding to photons passing through the left or right slits as shown in fig. \ref{fig:Concept}. 
%
%%%%%%%%%%%%%%%% Fig. 1 %%%%%%%%%%%%%%%%
%%%%%%%%%%%%%%%%%%%%%%%%%%%%%%%%%%%%%%%%
%
In this basis, the entangled state ideally generated by our setup is given by 
\begin{eqnarray}
|\Psi_{\textrm{slits}}\rangle=\frac{1}{\sqrt{2}}(|l\rangle_A|r\rangle_B+|r\rangle_A|l\rangle_B)
, 
\label{EntangledState}
\end{eqnarray}
where the suffixes A and B denote two photons found in different arms of our setup. 
In this section, we explain the setup in detail and present an experimental confirmation of the entanglement of our source by quantum tomography. 
%%%%%%%%%%%%%%%%%%%%%%%%%%%%%%%%%%%%%%%%
\subsection{Experimental setup\label{II-A}}
\begin{figure}
\includegraphics[width=0.678\linewidth,height=0.440\linewidth]{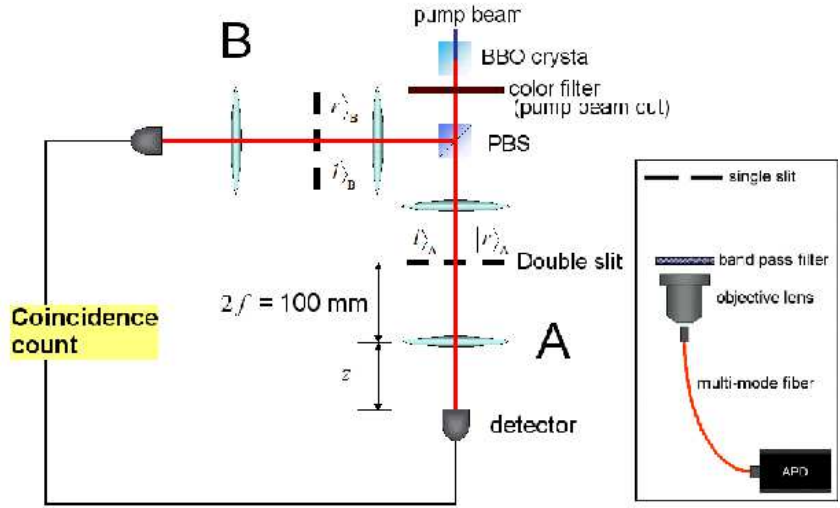}
\caption{\label{fig:ExpSys} Experimental setup. 
Note that basis states in the B arm are exchanged due to the reflection at the PBS. 
The detector system is composed of the parts shown in the inset. 
}
\end{figure}

%%%%%%%%%%%%%%%%%%%%%%%%%%%%%%%%%%%%%%%%
Figure \ref{fig:ExpSys} shows our experimental setup. 
%
%%%%%%%%%%%%%%%% Fig. 2 %%%%%%%%%%%%%%%%
%%%%%%%%%%%%%%%%%%%%%%%%%%%%%%%%%%%%%%%%
%
The entangled photons were generated in type II SPDC by a 405 nm pump beam from a 45 mW CW laser incident on a 5 mm-thick $\beta$-Barium Borate (BBO) crystal. 
The BBO crystal was placed in the collinear condition. 
The photons were separated into two different directions by a polarizing beam splitter (PBS). 
In each arm, a double-slit was placed in the focal plane of a lens in order to select a pair of transverse momenta of the photons. 
The slit width was 40$\mu$m and the distance between the slits was 150$\mu$m .
A lens of focal length $f=50$ mm was placed at a distance of $2f$ from the double-slit. 
A detector system was placed in a plane at a distance of $z$ from the lens. 
\par
The detector system consists of a single slit with a width of 40$\mu$m, a band pass filter (810 nm, band width of 10 nm), an objective lens to couple the photons into a multimode fiber, and a photon detector (Perkin-Elmer SPCM-AQR-14) as shown in the inset of fig. \ref{fig:ExpSys}. 
The coincidence counts between the detectors in the two arms were recorded. 
%%%%%%%%%%%%%%%%%%%%%%%%%%%%%%%%%%%%%%%%
\subsection{Measurement of Pauli operator averages in the focal and image planes\label{II-B}}
%%%%%%%%%%%%%%%%%%%%%%%%%%%%%%%%%%%%%%%%
\par
\begin{figure}
\includegraphics[width=1.0\linewidth,height=0.5\linewidth]{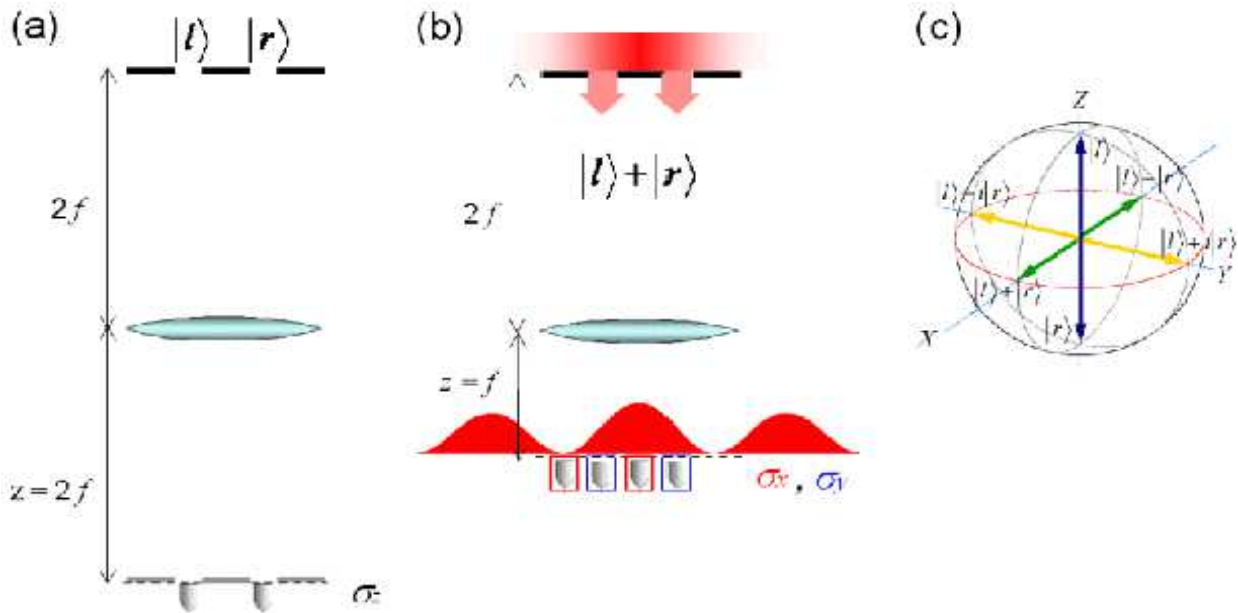}
\caption{\label{fig:slit_2f_lens} 
Measurements of Pauli operators. 
(a) 
$\sigma_z$ can be measured in the image plane. 
(b) 
Any linear combination of $\sigma_x$ and $\sigma_y$ can be measured in the focal plane. 
(c) 
Detectors in the image plane and the focal plane measure the states on the poles and the equator of the Bloch sphere. 
}
\end{figure}

The spatial qubits can be characterized in terms of the Pauli operators, $\sigma_x$, $\sigma_y$ and $\sigma_z$. 
In the $\{|l\rangle, |r\rangle\}$ basis, they are 
$\sigma_x = |l\rangle\langle r| + |r \rangle\langle l|$, 
$\sigma_y = -i\left(|l\rangle\langle r| - |r \rangle\langle l|\right)$, 
and $\sigma_z = |l\rangle\langle l| - |r \rangle\langle r|$. 
To measure $\sigma_z$, we have to distinguish the two slits. 
Since the image of the double-slit appears in the image plane at a distance of $z = 2f$ from the lens (fig. \ref{fig:slit_2f_lens}(a)), we can do this by placing the detectors at the positions corresponding to each slit in the image plane. 
%
%%%%%%%%%%%%%%%% Fig. 3 %%%%%%%%%%%%%%%%
%%%%%%%%%%%%%%%%%%%%%%%%%%%%%%%%%%%%%%%%
%
To measure $\sigma_x$ and $\sigma_y$, we have to obtain the phase information of the interference pattern of the double slits. 
Since the interference patterns are observed in the focal plane at a distance of $z = f$ from the lens (fig. \ref{fig:slit_2f_lens}(b)), we can do this by placing the detectors at appropriate positions in the focal plane. 
E. g. for a photon in the state $|l\rangle+|r\rangle$, an interference pattern appears as indicated in fig. \ref{fig:slit_2f_lens}(b). 
The $\sigma_x$ operator averages can then be estimated from the difference between the count rates at the center and at the first node of this interference pattern. 
%%%%%
These two detector positions thus correspond to projective measurements of the $\sigma_x$ eigenstates, $|l\rangle+|r\rangle$ and $|l\rangle-|r\rangle$. 
Likewise, a measurement of the $\sigma_y$ eigenstates can be realized by placing the detectors at the positions corresponding to $|l\rangle+i|r\rangle$ and $|l\rangle-i|r\rangle$. 
We can therefore realize the Pauli operator measurements by detecting photons at each of the six positions in the two planes corresponding to the eigenstates of $\sigma_x$, $\sigma_y$, and $\sigma_z$, as shown in fig. \ref{fig:slit_2f_lens}(c).
%%%%%%%%%%%%%%%%%%%%%%%%%%%%%%%%%%%%%%%%
\subsection{Results of quantum tomography\label{II-C}}
%%%%%%%%%%%%%%%%%%%%%%%%%%%%%%%%%%%%%%%%
%
%%%%%%%%%%%%%%%% Fig. 4 %%%%%%%%%%%%%%%%
\begin{figure}
\includegraphics[width=1.0\linewidth,height=0.64\linewidth]{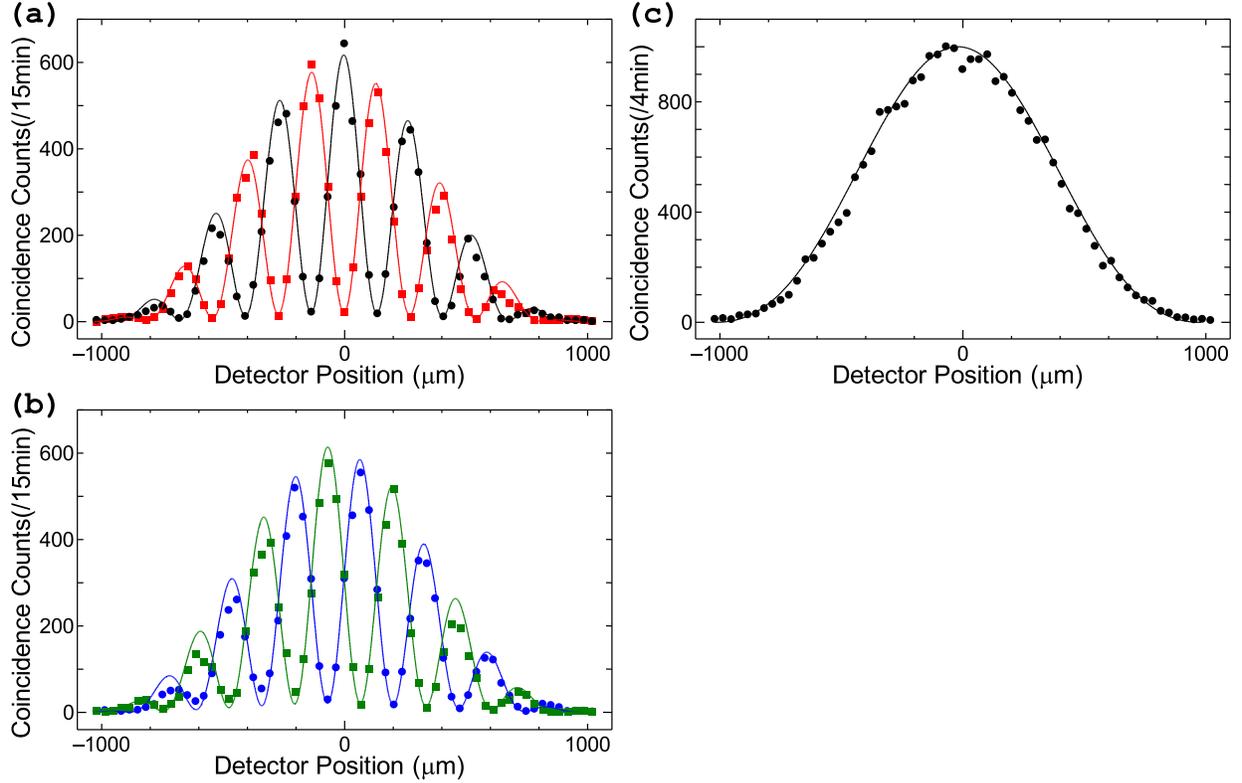}
\caption{\label{fig:ExpResults1} 
Coincidence count rates as a function of scanning detector position. 
The dots and squares show the experimental data and the solid curves are guides for the eyes. 
The detector positions at 0 and 135$\mu$m correspond to a $\sigma_x$ measurement in the scanning arm A. 
The positions $\pm$67$\mu$m correspond to a $\sigma_y$ measurement in the scanning arm A. 
(a)Interference correlation. The detector in arm B was set to the $\sigma_x$ eigenstates $|l\rangle+|r\rangle$ (dots) and $|l\rangle-|r\rangle$ (squares). 
(b) 
The detector in arm B was set to the $\sigma_y$ eigenstates $|l\rangle+i|r\rangle$ (dots) and $|l\rangle-i|r\rangle$ (squares). 
(c) Interference - which-path correlation. The detector in arm B was set to a position corresponding to one of the $\sigma_z$ eigenstates ($|l\rangle$). 
}
\end{figure}
%%%%%%%%%%%%%%%%%%%%%%%%%%%%%%%%%%%%%%%%
%
%
%%%%%%%%%%%%%%%% Fig. 5 %%%%%%%%%%%%%%%%
\begin{figure}
\includegraphics[width=1.0\linewidth,height=0.394\linewidth]{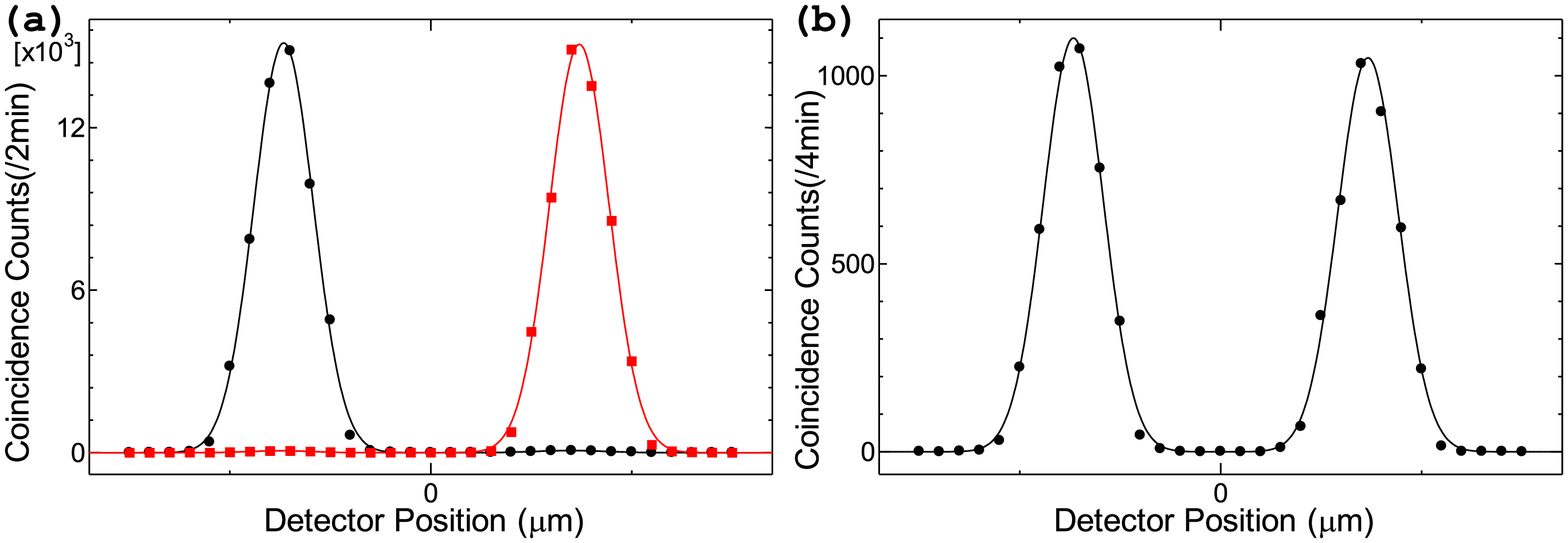}
\caption{\label{fig:ExpResults2} 
Coincidence count rates as a function of scanning detector position. 
The dots and squares show the experimental data and the solid curves are guides for the eyes. 
(a) Which-path correlation measurement. 
The detector in arm B was set to the $\sigma_z$ eigenstates $|l\rangle$ (dots) and $|r\rangle$ (squares). 
(b) One of the which-path - interference correlation measurements. 
The detector in arm B was set to a position corresponding to one of the $\sigma_x$ eigenstates ($|l\rangle+|r\rangle$). 
}
\end{figure}
%%%%%%%%%%%%%%%%%%%%%%%%%%%%%%%%%%%%%%%%
%
To measure the correlations between the spatial qubits, the detector in arm B was fixed at a position corresponding to one of the eigenstates of the Pauli operators. 
In arm A, the detector position $x$ was scanned in either the focal plane or the image plane. 
First, the scanning detector was placed in the focal plane. 
When the fixed detector was placed at the appropriate position for a measurement of a $\sigma_x$ or $\sigma_y$ eigenstate, a conditional interference pattern was observed. 
As the fixed detector position was moved, the interference pattern changed its phase without losing visibility. 
The interference patterns for the four eigenstates of $\sigma_x$ or $\sigma_y$ are shown in fig. \ref{fig:ExpResults1}(a) and (b). 
When the fixed detector was set at a position corresponding to a $\sigma_z$ eigenstate, the interference pattern disappeared as shown in fig. \ref{fig:ExpResults1}(c), that is, we see no correlation between $\sigma_z$ and $\sigma_x$/$\sigma_y$. 
Next, the scanning detector was placed in the image plane. 
The results are shown in fig. \ref{fig:ExpResults2}. 
The $\sigma_z$-$\sigma_z$ correlation was clearly observed, whereas no $\sigma_z$-$\sigma_x$ and $\sigma_z$-$\sigma_y$ correlation was found, as expected. 
\par
\begin{figure}
\includegraphics[width=0.78\linewidth,height=0.38\linewidth]{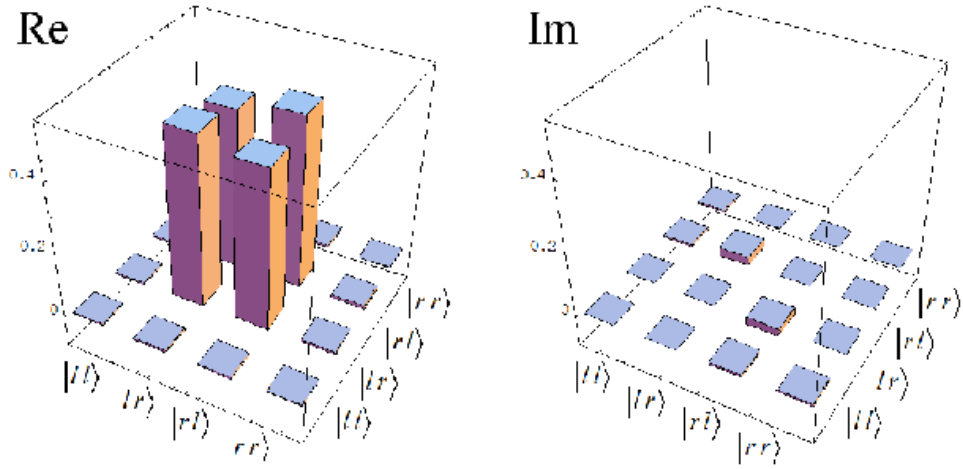}
\caption{\label{fig:DensityMatrix1} Real and imaginary part of the reconstructed density matrix. 
}
\end{figure}

The density matrix can be reconstructed by using the data extracted from the correlation experiments and listed in Table \ref{tbl:a1}(a). 
Specifically, the data for the $\sigma_x$ measurement in the scanning arm A was obtained from the scanning detector positions at 0 or 135$\mu$m from the center in fig. \ref{fig:ExpResults1}, and the data for the $\sigma_y$ measurement in the scanning arm A was obtained from the scanning detector positions at $\pm$67$\mu$m from the center in fig. \ref{fig:ExpResults1}. 
The data for the $\sigma_z$ measurement was taken from the centers of the slit images at $\pm 70\mu$m in fig. \ref{fig:ExpResults2}. 
%
%%%%%%% Table 1 %%%%%%%
\begin{table}
\caption{Coincidence counts. \label{tbl:a1}
The columns give the positions of the scanning arm A and the lines give the setting of the fixed arm B settings. 
}
(a) Raw data.
\hspace{2mm}
\begin{ruledtabular}
\begin{tabular}{|c|r|r|r|r|r|r|}%\hline
 & $|l\rangle$ & $|r\rangle$ & $|l\rangle+|r\rangle$ & $|l\rangle-|r\rangle$ & $|l\rangle+i|r\rangle$ & $|l\rangle-i|r\rangle$ \\\hline
$|l\rangle$ & 77 & 14838 & 995 & 999 & 957 & 967 \\\hline
$|r\rangle$ & 14885 & 66 & 993 & 888 & 953 & 970 \\\hline
$|l\rangle+|r\rangle$ & 1032 & 1071 & 643 & 22 & 340 & 288 \\\hline
$|l\rangle-|r\rangle$ & 1050 & 986 & 22 & 595 & 290 & 313 \\\hline
$|l\rangle+i|r\rangle$ & 1063 & 1053 & 309 & 308 & 554 & 29 \\\hline
$|l\rangle-i|r\rangle$ & 1008 & 1049 & 320 & 276 & 17 & 576 \\%\hline
\end{tabular}
\end{ruledtabular}
\vspace{5mm}
(b) Normalized data 
\begin{ruledtabular}
\begin{tabular}{|c|r|r|r|r|r|r|}%\hline
 & $|l\rangle$ & $|r\rangle$ & $|l\rangle+|r\rangle$ & $|l\rangle-|r\rangle$ & $|l\rangle+i|r\rangle$ & $|l\rangle-i|r\rangle$ \\\hline
$|l\rangle$ & 0.003 & 0.497 & 0.253 & 0.262 & 0.252 & 0.248 \\\hline
$|r\rangle$ & 0.498 & 0.002 & 0.252 & 0.233 & 0.251 & 0.249 \\\hline
$|l\rangle+|r\rangle$ & 0.245 & 0.255 & 0.486 & 0.017 & 0.275 & 0.227 \\\hline
$|l\rangle-|r\rangle$ & 0.258 & 0.242 & 0.017 & 0.480 & 0.243 & 0.255 \\\hline
$|l\rangle+i|r\rangle$ & 0.258 & 0.256 & 0.254 & 0.262 & 0.484 & 0.025 \\\hline
$|l\rangle-i|r\rangle$ & 0.238 & 0.248 & 0.256 & 0.228 & 0.014 & 0.477 \\%\hline
\end{tabular}
\end{ruledtabular}
\end{table}
%%%%%%%%%%%%%%%%%%%%%%%
%
To obtain the probabilities, it is necessary to normalize the results. 
Since the coincidence count rates decrease as the detectors move away from the optical axis, it is also necessary to compensate the resulting detection efficiency differences between the $|l\rangle+|r\rangle$ and $|l\rangle-|r\rangle$ states using the theoretically expected ratios between the corresponding peaks in fig. \ref{fig:ExpResults1}(a). 
The normalized results are shown in table \ref{tbl:a1}(b). 
From these probabilities, we can reconstruct the density matrix. 
In the $\{|ll\rangle,|lr\rangle,|rl\rangle,|rr\rangle\}$ basis, it is given by 
\begin{eqnarray}
\hat{\rho}_{AB}&=&
\left(\begin{array}{c c c c}
0.003 & -0.005-0.007i & -0.006+0.000i & 0.002-0.006i\\
-0.005+0.007i & 0.498 & 0.463-0.024i & 0.009+0.001i\\
-0.006 0.000i & 0.463+0.024i & 0.497 & 0.008-0.007i\\
0.002+0.006i & 0.009-0.001i & 0.008+0.007i & 0.002
\end{array}\right)
. 
\label{NumericalDensityMatrix1}
\end{eqnarray}
Figure \ref{fig:DensityMatrix1} shows an illustration of the density matrix elements. 
The fidelity of the ideal state given in eq. (\ref{EntangledState}) is
$F = \langle\Psi_{\textrm{slits}}|\hat{\rho}_{AB}|\Psi_{\textrm{slits}}\rangle=$0.961, indicating that we have successfully generated an entangled state close to the theoretically expected one. 
%
%%%%%%%%%%%%%%%% Fig. 6 %%%%%%%%%%%%%%%%
%%%%%%%%%%%%%%%%%%%%%%%%%%%%%%%%%%%%%%%%
%
%
%
%%%%%%%%%%%%%%%%%%%%%%%%%%%%%%%%%%%%%%%%
\section{Measurement between the focal and image planes\label{Sec3}}
\begin{figure}
\includegraphics[width=0.408\linewidth,height=0.216\linewidth]{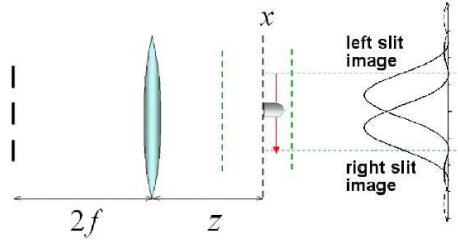}
\caption{\label{fig:MediumPlane} 
Simultaneous measurement of interference and which-path information realized by placing the detector between the focal and image planes. 
}
\end{figure}

%%%%%%%%%%%%%%%%%%%%%%%%%%%%%%%%%%%%%%%%
The measurement of an arbitrary superposition state of the spatial qubit may play an important role in the application of the spatial qubit entanglement. 
In this section, we show that such measurements can be realized by placing the detector between the focal and image planes, as shown in fig. \ref{fig:MediumPlane}. 
The positive operator valued measure describing the measurement in this intermediate plane is derived and its application to state preparation and quantum tomography is discussed. 
%
%%%%%%%%%%%%%%%% Fig. 7 %%%%%%%%%%%%%%%%
%%%%%%%%%%%%%%%%%%%%%%%%%%%%%%%%%%%%%%%%
%
%%%%%%%%%%%%%%%%%%%%%%%%%%%%%%%%%%%%%%%%
\subsection{Positive operator valued measure in the double-slit basis\label{III-A}}
%%%%%%%%%%%%%%%%%%%%%%%%%%%%%%%%%%%%%%%%
The quantum states of the photons at the double-slits can be expressed in terms of a transverse spatial wavefunction $\varphi(0;x)$. 
If we place a lens of focal length $f$ at a distance $L$ from the slits, 
then the transverse wavefunction $\varphi(L;x)$ in the plane in front of the lens can be calculated using the Fresnel-Kirchhoff diffraction integral
%%%%%%%%%%%%%%%%%%%%%%%%%%%%%%%%%%%%%%%%
\begin{eqnarray}
\varphi(L;x)
&=&
\sqrt{\frac{i}{\lambda L}} e^{i\frac{2\pi L}{\lambda}} e^{i\frac{\pi}{\lambda{L}}x^2}
\int_{S_0} \varphi(0;x') e^{i\frac{\pi}{\lambda L}{x'}^2} e^{-i\frac{2\pi}{\lambda L}xx'} dx'
. 
\end{eqnarray}
%%%%%%%%%%%%%%%%%%%%%%%%%%%%%%%%%%%%%%%%
where the integral is performed over the one-dimensional slit plane. 
The diffraction at the lens modifies the free space propagation so that the wavefunction at the distance $z$ from the lens corresponds to the wavefunction at an effective propagation length of $R = (Lf+zf-Lz)/(z-f)$ from the double-slits, reduced in size by a factor of $(z-f)/f$. 
Up to an arbitrary phase factor, the wavefunction at $z$ is then given by 
%Similarly, the wavefunction at a distance $z$ from the lens is calculated as 
%%%%%%%%%%%%%%%%%%%%%%%%%%%%%%%%%%%%%%%%
\begin{eqnarray}
\varphi(L+z;x)
&=&
\sqrt{\frac{f}{\lambda R(z-f)}}
e^{-i\frac{\pi}{\lambda R}\frac{L-f}{z-f}x^2}
\int_{S_0}
\varphi(0;x')
e^{-i\frac{\pi}{\lambda R}{x'}^2}
e^{-i\frac{2\pi}{\lambda R}\frac{f}{z-f}xx'}
dx'
%%%%%%%%%%%%%%%%%%%%%%%%%%%%%%%%%%%%%%%%
. 
\end{eqnarray}
%%%%%%%%%%%%%%%%%%%%%%%%%%%%%%%%%%%%%%%%
%
\par
The wavefunction of a slit of width $a$ at a distance of $r_n$ from the optical axis is given by 
%%%%%%%%%%%%%%%%%%%%%%%%%%%%%%%%%%%%%%%%
\begin{eqnarray}
\varphi_n(L+z;x)
&\equiv&
%%%%%%%%%%%%%%%%%%%%%%%%%%%%%%%%%%%%%%%%
\sqrt{\frac{f}{\lambda Ra(z-f)}}
e^{-i\frac{\pi}{\lambda R}\frac{L-f}{z-f}x^2}
\int_{-a/2}^{a/2}
e^{-i\frac{\pi}{\lambda R}{\left({x'}+r_n\right)}^2}
e^{-i\frac{2\pi}{\lambda R}\frac{f}{z-f}x\left({x'}+r_n\right)}
dx'
, 
\label{WavFunc}
\end{eqnarray}
%%%%%%%%%%%%%%%%%%%%%%%%%%%%%%%%%%%%%%%%
where we have assumed that the wavefunction in the slit is uniform. 
If the detector plane is far enough from the image plane, so that $R > a^2/\lambda$, eq. (\ref{WavFunc}) can be simplified to the conventional slit diffraction pattern given by the sinc function $\textrm{sinc}(x)=\sin x/x$, 
%%%%%%%%%%%%%%%%%%%%%%%%%%%%%%%%%%%%%%%%
\begin{eqnarray}
\varphi_n(L+z;x)
&=&
\sqrt{\frac{K}{\pi}}
\exp\left(-i\frac{2 r_n}{a}K x\right)\textrm{sinc}\left[K\left(x+\frac{z-f}{f}r_n\right)\right]
, 
\label{ApproxWavFunc}
\end{eqnarray}
%%%%%%%%%%%%%%%%%%%%%%%%%%%%%%%%%%%%%%%%
where $K = \pi a f/\lambda R(z-f)$ defines the scale of the single slit diffraction pattern. 
%%%%%%%%%%%%%%%%%%%%%%%%%%%%%%%%%%%%%%%%
\par
For the double-slit system, $r_l=-d/2$ and $r_r=d/2$. 
Each basis state of the qubit system thus corresponds to a well-defined wavefunction in the plane at $z$ from the lens. 
These two wavefunctions define the two dimensional Hilbert space of the qubit in the measurement plane. 
It is then possible to express the effect of a measurement of $x$ in this plane by a projection onto a non-normalized state in the Hilbert space of the qubit, 
\begin{eqnarray}
|m(x)\rangle&=&
\varphi_l^*(L+z;x)|l\rangle+\varphi_r^*(L+z;x)|r\rangle
. 
\label{MeasurementBasis}
\end{eqnarray}
We can therefore identify each measurement result $x$ with a specific projection state of the spatial qubit, corresponding to a point on the Bloch sphere. 
The positive operator valued measure describing the qubit measurement is given by the measurement operators
\begin{eqnarray}
\hat{M}(x)\equiv|m(x)\rangle\langle m(x)|
. 
\label{MeasurementOperator}
\end{eqnarray}
For a qubit state given by a density operator $\hat{\rho}$, the probability of photon detection at the position $x$ is 
\begin{eqnarray}
P(x)=\textrm{Tr}[\hat{M}\hat{\rho}]
, 
\end{eqnarray}
where the normalization of the wavefunctions $\varphi_l(L+z;x)$ and $\varphi_r(L+z;x)$ ensures that the set of operators $\{\hat{M}(x)\}$ satisfies the completeness relation
\begin{eqnarray}
\int_{-\infty}^{\infty}\hat{M}(x)dx = 1
. 
\label{Completeness}
\end{eqnarray}
%%%%%%%%%%%%%%%%%%%%%%%%%%%%%%%%%%%%%%%%
\par
We can easily extend this method to the analysis of multi-dimensional qudits. 
All we have to do is increase of the number of slits to $N$.
The projection state of eq. (\ref{MeasurementBasis}) is then given by a superposition of all slit states, 
\begin{eqnarray}
|m^{(N)}(x)\rangle
=
\sum_{n=1}^N
\varphi_n^*(L+z;x)|n\rangle
, 
\label{MeasurementBasesMulti}
\end{eqnarray}
where $\varphi_n(L+z;x)$ represents the wavefunction of a photon originating from the $n$th slit, as given in eq. (\ref{ApproxWavFunc}). 
%%%%%%%%%%%%%%%%%%%%%%%%%%%%%%%%%%%%%%%%
\subsection{Accessibility of arbitrary quantum states\label{III-B}}
%%%%%%%%%%%%%%%%%%%%%%%%%%%%%%%%%%%%%%%%
\par
\begin{figure}
\includegraphics[width=1.0\linewidth,height=0.346\linewidth]{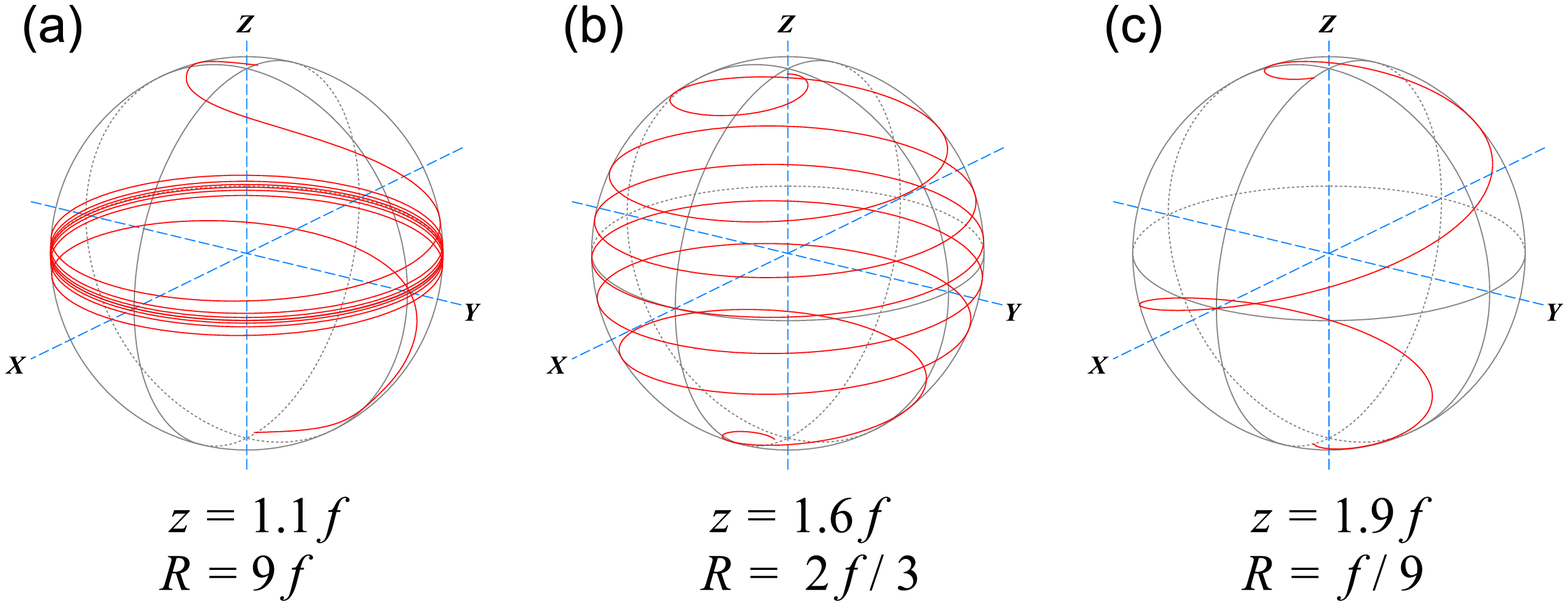}
\caption{\label{fig:BlochSphere} 
Measurement states $|m(x)\rangle$ on the Bloch sphere. 
The calculation was performed with the distance $z$ of the detector from the lens fixed, while the transverse position $x$ was varied until the intensity from one of the slits went down to almost zero. 
(a)-(c) show the trajectories of the Bloch vectors for the respective distance $z$ of the detector from the lens. 
}
\end{figure}

In our setup, $L$ is equal to $2f$. 
In this case, the effective propagation length is $R=f(2f-z)/(z-f)$, so we can access any value of $R$ between $R=0$ (image plane) and $R\to \infty$ (focal plane) by placing the detectors at a distance $z$ from the lens between $z=f$ and z=2$f$. 
For each detector setting, the projection states $|m(x)\rangle$ given by eq. (\ref{MeasurementBasis}) trace out a trajectory on the Bloch sphere as $x$ is scanned. 
The trajectories for three specific detector positions $z$ are shown in fig. \ref{fig:BlochSphere}. 
In each case, the Bloch vector moves from the upper hemisphere to the lower hemisphere rotating around the $\sigma_z$ axis as the detector position $x$ is scanned. 
For $z$ close to $f$(fig. \ref{fig:BlochSphere}(a)), the Bloch vectors are concentrated around the equator. 
They spread out as $z$ increases and accumulate at the poles as $z$ approaches 2$f$(fig. \ref{fig:BlochSphere}(c)). 
In the intermediate case shown in fig. \ref{fig:BlochSphere}(b), the Bloch vectors are almost equally distributed over the Bloch sphere, indicating that any state on the Bloch sphere can be measured by an appropriate choice of detector position ($z$, $x$). 
%
%%%%%%%%%%%%%%%% Fig. 8 %%%%%%%%%%%%%%%%
%%%%%%%%%%%%%%%%%%%%%%%%%%%%%%%%%%%%%%%%
%
%%%%%%%%%%%%%%%%%%%%%%%%%%%%%%%%%%%%%%%%
\par
Using our source of entangled qubits, we can prepare an arbitrary state of one spatial qubit by post-selecting a specific measurement outcome of the other qubit. 
For the ideal state of eq. (\ref{EntangledState}), the non-normalized state prepared in A by post-selecting a measurement result of $x$ in B is 
\begin{eqnarray}
|\Psi_{\textrm{prep}}\rangle_{A}
={}_B\langle m(x_B)|\Psi_\textrm{slit}\rangle_{AB}
=\frac{1}{\sqrt{2}}\left(
\varphi_r(L+z;x_B)|l\rangle + \varphi_l(L+z;x_B)|r\rangle
\right)
.
\label{PreparedState}
\end{eqnarray}
%%%%%%%%%%%%%%%%%%%%%%%%%%%%%%%%%%%%%%%%
The Bloch vector of this state is the mirror image of the Bloch vector of $|m(x)\rangle$ at the $X-Y$ plane. 
The trajectory of states that can be prepared in A by measurements in B are therefore equivalent to the trajectories shown in fig. \ref{fig:BlochSphere}. 
%%%%%%%%%%%%%%%%%%%%%%%%%%%%%%%%%%%%%%%%
\subsection{Spatial patterns of density matrix elements\label{III-C}}
%%%%%%%%%%%%%%%%%%%%%%%%%%%%%%%%%%%%%%%%
From the count rate pattern obtained by scanning the detector position $x$ in a plane between the focal and image planes, we can know the state of the spatial qubit. 
In the double-slit basis, the density matrix of the spatial qubit at the double-slit is given by 
%%%%%%%%%%%%%%%%%%%%%%%%%%%%%%%%%%%%%%%%
\begin{eqnarray}
\hat{\rho}&=&
\left(\begin{array}{c c}
\rho_{ll} & \rho_{lr}\\
\rho_{rl} & \rho_{rr}
\end{array}\right)
. 
\label{DensityMatrix1}
\end{eqnarray}
%%%%%%%%%%%%%%%%%%%%%%%%%%%%%%%%%%%%%%%%
For a detector position $z$, the positive operator valued measure of eq. (\ref{MeasurementOperator}) gives the probability of detection at a point $x$ as 
%%%%%%%%%%%%%%%%%%%%%%%%%%%%%%%%%%%%%%%%
\begin{eqnarray}
P(x)
&=&\textrm{Tr}\left[\hat{\rho}\hat{M}\right]\nonumber
\\
&=&
\rho_{ll}\left|\varphi_l(L+z;x)\right|^2
+\rho_{lr}\;\varphi_l^*(L+z;x)\varphi_r(L+z;x)
\nonumber
\\&{}&
+\rho_{rl}\;\varphi_l(L+z;x)\varphi_r^*(L+z;x)
+\rho_{rr}\left|\varphi_r(L+z;x)\right|^2
. 
\label{TraceRhoM1}
\end{eqnarray}
%%%%%%%%%%%%%%%%%%%%%%%%%%%%%%%%%%%%%%%%
Each density matrix element is thus connected to a unique pattern defined by $\varphi_l$ and $\varphi_r$. 
Using eq. (\ref{ApproxWavFunc}), we obtain 
\begin{eqnarray}
P(x)&=&
\rho_{ll}\frac{K}{\pi}\left|\textrm{sinc}\left(Kx-\frac{\Delta\phi}{2}\right)\right|^2
\nonumber\\&{}&
+\rho_{lr}\frac{K}{\pi}\exp\left(-i\frac{2d}{a}Kx\right)
\textrm{sinc}\left(Kx-\frac{\Delta\phi}{2}\right)
\textrm{sinc}\left(Kx+\frac{\Delta\phi}{2}\right)
\nonumber\\&{}&
+\rho_{rl}\frac{K}{\pi}\exp\left(i\frac{2d}{a}Kx\right)
\textrm{sinc}\left(Kx-\frac{\Delta\phi}{2}\right)
\textrm{sinc}\left(Kx+\frac{\Delta\phi}{2}\right)
\nonumber
\\&{}&
+\rho_{rr}\frac{K}{\pi}\left|\textrm{sinc}\left(Kx+\frac{\Delta\phi}{2}\right)\right|^2
, 
\label{ApproxProb1}
\end{eqnarray}
%%%%%%%%%%%%%%%%%%%%%%%%%%%%%%%%%%%%%%%%
where 
$\Delta\phi = (z-f) Kd/ f$ describes the spatial displacement of the diffraction patterns of the two slits, that is, $\Delta\phi$ is proportional to the ratio of the separation between the slit images and the width of the diffraction pattern of a slit. 
The closer the detection plane comes to the focal plane, the smaller the separation between the slit images becomes. 
%%%%%%%%%%%%%%%%%%%%%%%%%%%%%%%%%%%%%%%%
\par
In general, the detection probability $P(x)$ is a linear function of the four density matrix elements. 
Therefore, it is in principle always possible to invert and solve the equation for the elements, so that the density matrix elements can be determined from specific integrals of $P(x)$. 
In the case of the qubit in eq. (\ref{ApproxProb1}), the complete density matrix can be determined by only two coefficients, the difference between the diagonal elements $\rho_{ll}-\rho_{rr}$ and the complex off-diagonal element $\rho_{lr}$. 
In $P(x)$, $\rho_{ll}-\rho_{rr}$ appears as a difference between the squared sinc functions describing the diffraction patterns of the two slits and $\rho_{lr}$ appears as an oscillating interference pattern with an envelope given by the product of the two sinc functions. 
If the oscillation is sufficiently fast ($d \gg a$), the inversion can be performed by integrating the product of $P(x)$ and the patterns corresponding to $\rho_{ll}-\rho_{rr}$ and $\rho_{lr}$, respectively. 
The result of the inversion is then given by 
%%%%%%%%%%%%%%%%%%%%%%%%%%%%%%%%%%%%%%%%
\begin{eqnarray}
%%%%%%%%%%%%%%%%%%%%
\rho_{ll}-\rho_{rr}
&=&
\frac{3}{2(1-\beta)}
\int
\left[
\textrm{sinc}^2\left(Kx-\frac{\Delta\phi}{2}\right)
-\textrm{sinc}^2\left(Kx+\frac{\Delta\phi}{2}\right)
\right]
\;P(x)\;dx
\nonumber
%\label{IntPositionProb}
%%%%%%%%%%%%%%%%%%%%
\\
\rho_{lr}
&=&
\frac{3}{2\beta}
\int
\exp\left(i\frac{2d}{a}Kx\right)
\textrm{sinc}\left(Kx-\frac{\Delta\phi}{2}\right)
\textrm{sinc}\left(Kx+\frac{\Delta\phi}{2}\right)
\;P(x)\;dx
, 
\nonumber \\[0.3cm]
&& \mbox{where} \hspace{0.5cm} \beta=\frac{3}{2 \Delta\phi^2}(1-\textrm{sinc}(2\Delta\phi)).
\label{IntExpProb}
\end{eqnarray}
%%%%%%%%%%%%%%%%%%%%%%%%%%%%%%%%%%%%%%%%
Since $\rho_{ll}+\rho_{rr} = 1$, these two patterns completely define the density matrix of the qubit. 
For higher dimensional qudit systems, similar relations can be derived for all $N^2$ density matrix elements of an $N$-slit system. 
Thus the density matrix of the spatial qudits can be directly identified with the corresponding spatial patterns observed in the measurement. 
%
%
%%%%%%%%%%%%%%%%%%%%%%%%%%%%%%%%%%%%%%%%
\section{Experimental results\label{Sec4}}
%%%%%%%%%%%%%%%%%%%%%%%%%%%%%%%%%%%%%%%%
In this section, we present the experimental results of the measurements with the detector placed between the focal and image planes. 
Conditional measurements of the pattern $P(x)$ in arm A were performed with a fixed detector position $x$ in arm B. 
From this data, we can determine the conditional density matrices of the states in A. 
The density matrix of the complete two qubit system is then reconstructed from all 6 conditional measurements and the result is compared with that obtained in section \ref{Sec2}. 
%%%%%%%%%%%%%%%%%%%%%%%%%%%%%%%%%%%%%%%%
\subsection{State preparation and conditional density matrices\label{IV-A}}
\begin{figure}
\includegraphics[width=0.27\linewidth,height=0.261\linewidth]{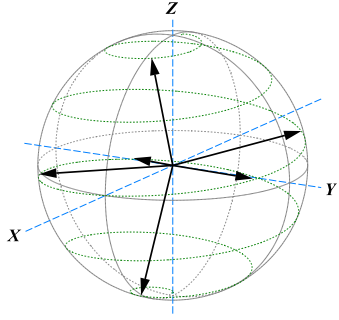}
\caption{\label{fig:octahedron} 
Six Bloch vectors corresponding to the detector positions in the fixed arm B. 
}
\end{figure}

\begin{figure}
\includegraphics[width=1.0\linewidth,height=0.47\linewidth]{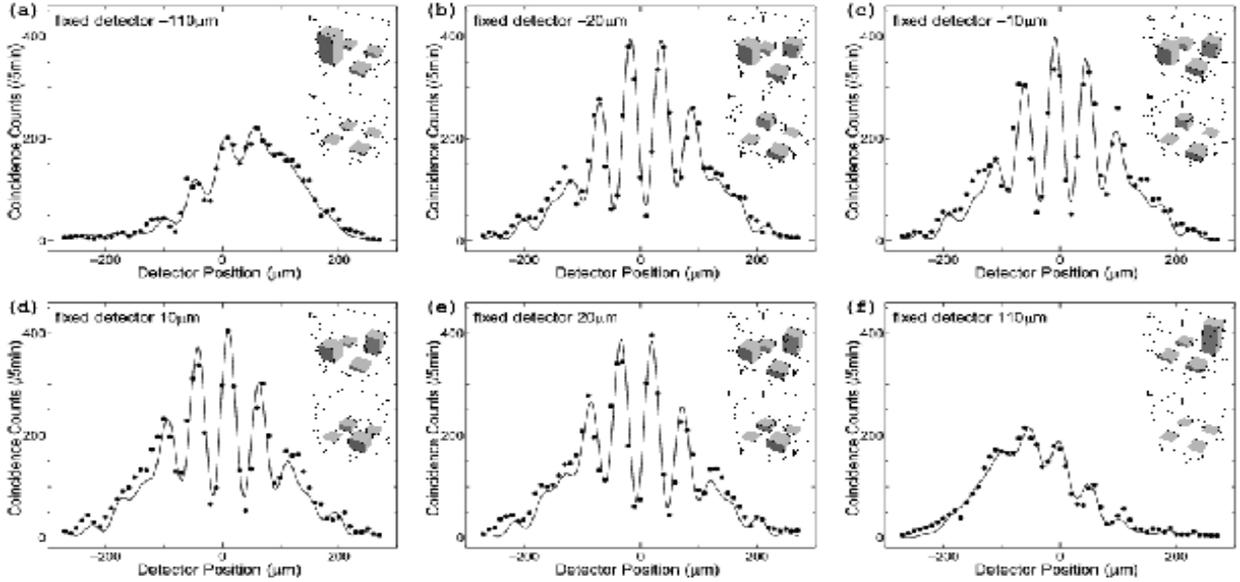}
\caption{\label{fig:ConditinalScanningData} Coincidence counts data. 
Each measurement was obtained by placing the fixed detector in a different position. 
The insets show the corresponding conditional density matrices.
}
\end{figure}

%%%%%%%%%%%%%%%%%%%%%%%%%%%%%%%%%%%%%%%%
The experimental setup was as described in section \ref{Sec2}, except that the slit width in the detector system was 20$\mu$m. 
The distance $z$ was set to $1.8 f$ in each arm. 
Six different detector positions were chosen for arm B such that the corresponding Bloch vectors are close to a regular octahedron, as shown in fig. \ref{fig:octahedron}. 
The detector in A was then scanned to obtain the coincidence count data shown in fig. \ref{fig:ConditinalScanningData}. 
The distributions measured in each scan correspond to a one qubit density matrix, where the diagonal elements determine the bias between the right and left side of the distributions and the off-diagonal elements determine the visibility and the phase of the interference patterns. 
%
%%%%%%%%%%%%%%%% Fig. 9 %%%%%%%%%%%%%%%%
%%%%%%%%%%%%%%%%%%%%%%%%%%%%%%%%%%%%%%%%
%
%
%%%%%%%%%%%%%%%% Fig. 10 %%%%%%%%%%%%%%%
%%%%%%%%%%%%%%%%%%%%%%%%%%%%%%%%%%%%%%%%
%
%%%%%%%%%%%%%%%%%%%%%%%%%%%%%%%%%%%%%%%%
\par
We can understand the results of the conditional measurements from the viewpoint of quantum state preparation as explained in section \ref{III-B}. 
In our experiments, the detector position in the fixed arm B selects the prepared state in the scanning arm A. 
We can check how well the conditional state in arm A corresponds to the expected one described by eq. (\ref{PreparedState}) by reconstructing the conditional density matrix and determining the fidelity. 
To reconstruct the density matrix and to determine the actual value of the experimental parameter $L$ at the same time, we performed a least square fit based on eq. (\ref{TraceRhoM1}). 
The result of the fits and the corresponding density matrices are shown in fig. \ref{fig:ConditinalScanningData}. 
The fidelities of the expected prepared states given by eq. (\ref{PreparedState}) are given in table \ref{tbl:fidelity}. 
The results show that the states prepared in A by measurements in B are close to the intended ones. 
%
%%%%%%% Table 2 %%%%%%%
\begin{table}
\caption{\label{tbl:fidelity} Fidelities of expected prepared states and experimentally obtained states in fig. \ref{fig:ConditinalScanningData} }
\begin{ruledtabular}
\begin{tabular}{|c|c|c|c|c|c|c|}%\hline
case & (a) & (b) & (c) & (d) & (e) & (f) \\\hline
fidelity & 0.887 & 0.861 & 0.841 & 0.841 & 0.871 & 0.912 \\%\hline
\end{tabular}
\end{ruledtabular}
\end{table}
%%%%%%%%%%%%%%%%%%%%%%%
%%%%%%%%%%%%%%%%%%%%%%%%%%%%%%%%%%%%%%%%
\subsection{Reconstruction of the density matrix from conditional scans\label{IV-B}}
%%%%%%%%%%%%%%%%%%%%%%%%%%%%%%%%%%%%%%%%
The non-normalized conditional density matrices $\hat{\rho}_A(x_i)$ obtained by setting the detector in arm B to $x_i$ depend on the density matrix $\hat{\rho}_{AB}$ of the entangled qubit according to
\begin{eqnarray}
\hat{\rho}_A(x_i)={}_B\langle m(x_i)|\hat{\rho}_{AB}|m(x_i)\rangle_B
. 
\end{eqnarray}
In principle, four different measurement points $x_i$ are enough to reconstruct the density matrix in B. 
The entangled density matrix of A and B can be reconstructed from the conditional density matrices $\hat{\rho}_A(x_i)$ with the same linear coefficients used to reconstruct a density matrix in B from probabilities $P(x_i)$.
Specifically, the reconstruction of $\hat\rho_B$ from $P(x_i)$ and the reconstruction of $\hat{\rho}_{AB}$ from $\hat\rho_A(x_i)$ are related by a set of reconstruction operators $\hat{\Lambda}_i$ with 
\begin{eqnarray}
\hat{\rho}_B &=& \sum_i P(x_i) \hat\Lambda_i
\\
\hat{\rho}_{AB} &=& \sum_i \hat\rho_A(x_i) \otimes\hat\Lambda_i
. 
\end{eqnarray}
Since we have chosen 6 measurement points in B, we have sufficient information to reconstruct the complete two qubit density matrix $\hat{\rho}_{AB}$ from the conditional density matrices $\hat{\rho}_A(x_i)$. 
The result in the $\{|ll\rangle,|lr\rangle,|rl\rangle,|rr\rangle\}$ basis reads
\begin{eqnarray}
\hat{\rho}_{AB}&=&
\left(\begin{array}{c c c c}
 0.008 & 0.008-0.012i & 0.015+0.021i & -0.018-0.001i\\
 0.008+0.012i & 0.485 & 0.347-0.038i & 0.002-0.027i\\
 0.015-0.021i & 0.347+0.038i & 0.469 & 0.008+0.005i\\
-0.018+0.001i & 0.002+0.027i & 0.008-0.0050i & 0.038\\
\end{array}\right)
. 
\label{NumericalDensityMatrix2}
\end{eqnarray}
The fidelity of the ideal state given in eq. (\ref{EntangledState}) is $F = \langle\Psi_{\textrm{slits}}|\hat{\rho}_{AB}|\Psi_{\textrm{slits}}\rangle$ = 0.824. 
%%%%%%%%%%%%%%%%%%%%%%%%%%%%%%%%%%%%%%%%
\subsection{Discussion of the results\label{IV-C}}
%%%%%%%%%%%%%%%%%%%%%%%%%%%%%%%%%%%%%%%%
At present, the agreement with the theoretical prediction 1 is not as good as that of the Pauli operator measurements described in section \ref{Sec2}. 
It should therefore be possible to improve the experiment by identifying the sources of the additional errors. 
Since there are some discrepancies between the experimental data and the theoretical fit in fig. \ref{fig:ConditinalScanningData}, one problem may be that the experimental parameters used in the theory need to be adjusted. 
Moreover, the visibilities may be underestimated by the fit as indicated by data points above and below the maxima and minima of the interference patterns. 
An increase in the spatial resolution of the measurement might solve this problem. 
\par
The main merit of our method is that we can see both the which-path and the interference information in a single scan. 
Our experimental results clearly confirm that the coincidence count distribution contains all the data necessary to reconstruct the density matrix. 
This kind of single scan tomography thus provides a direct characterization of spatial qudits in terms of their actual physical properties in space. 
%
%
%%%%%%%%%%%%%%%%%%%%%%%%%%%%%%%%%%%%%%%%
\section{Conclusions\label{Sec5}}
%%%%%%%%%%%%%%%%%%%%%%%%%%%%%%%%%%%%%%%%
We have presented a thorough experimental investigation of the entanglement between a pair of spatial qubits generated by passing down-converted photon pairs through double-slits. 
Pauli operator measurements indicate that we have achieved a fidelity of $F = 0.961$ for the maximally entangled two qubit states. 
We have then explored the possibility of accessing arbitrary superposition states by measurements between the focal and image planes. 
We can use the entanglement to prepare arbitrary states in arm A by measurements in arm B. 
The results have been verified by single scan tomography, where we identified each density matrix element with its distinct pattern in the measurement distribution. 
Fidelities between 0.841 and 0.912 have been obtained for the conditionally prepared states. 
By combining the conditional results, we obtained a two qubit density matrix with a fidelity of 0.824. 
The decrease in the fidelity compared to the Pauli operator measurements suggests that further experimental improvements may be possible. 
%%%%%%%%%%%%%%%%%%%%
\par
The method introduced here allows us to access the full Hilbert space of the spatial qubits. 
It may therefore have applications in the realization of quantum information protocols using double-slit qubits.
Moreover, the extension of our method to $N$-slit qudits is straightforward, opening the way to quantum operations in higher dimensional Hilbert spaces. 
%
%
%%%%%%%%%%%%%%%%%%%%%%%%%%%%%%%%%%%%%%%%
\begin{acknowledgments}
%%%%%%%%%%%%%%%%%%%%%%%%%%%%%%%%%%%%%%%%
We are grateful to Takao Hirama for his devotion to work on getting our experiment started. 
Part of this work has been suppoorted the Grant-in-Aid program of the Japanese Society for the Promotion of Science. 
\end{acknowledgments}
%
%
%\newpage %Just because of unusual number of tables stacked at end
%\bibliography{bib003}% bib003.bib

\begin{thebibliography}{xyz00}

\bibitem{Ou1988}
Z. Y. Ou and L. Mandel, 
Phys. Rev. Lett., {\bf 61}, 50, (1988).

\bibitem{Kwiat1995}
P. G. Kwiat, K. Mattle, H. Weinfurter, A. Zeilinger, A. V. Sergienko, and Y. Shih, 
Phys. Rev. Lett., {\bf 75}, 4337, (1995).

\bibitem{Franson1989}
J. D. Franson, 
Phys. Rev. Lett., {\bf 62}, 2205, (1989).

\bibitem{Strekalov1995}
D.V. Strekalov, A.V. Sergienko, D.N. Klyshko, and Y.H. Shih, 
Phys. Rev. Lett., {\bf 74}, 3600, (1995).

\bibitem{Howell2004}
J. C. Howell, R. S. Bennink, S. J. Bentley, and R. W. Boyd, 
Phys. Rev. Lett., {\bf 92}, 210403, (2004).

\bibitem{D'Angelo2004}
M. D'Angelo, Y. H. Kim, S. P. Kulik, and Y. Shih, 
Phys. Rev. Lett., {\bf 92}, 233601, (2004).

\bibitem{Kaszlikowski2000}
D. Kaszlikowski, P. Gnaci\'{n}ski, M. \.{Z}ukowski, W. Miklaszewski, and A. Zeilinger, 
Phys. Rev. Lett., {\bf 85}, 4418, (2000).

\bibitem{Joo2007}
J. Joo, P.L. Knight, J.L. O'Brien, and T. Rudolph, 
Phys. Rev. A, {\bf 76}, 052326, (2007).

\bibitem{Lanyon2008a}
B.P. Lanyon, M. Barbieri, M.P. Almeida, T. Jennewein, T.C. Ralph, K. J. Resch, G.J. Pryde, J.L. O'Brien, A. Gilchrist, and A. G. White, 
e-print, arXiv:0804.0272v1 (2008).

\bibitem{Mair2001}
A. Mair, A. Vaziri, G. Weihs, and A. Zeilinger, 
Nature, {\bf 412}, 313, (2001).

\bibitem{Vaziri2003}
A. Vaziri, G. Weihs, and A. Zeilinger, 
Phys. Rev. Lett., {\bf 89}, 240401, (2002).

\bibitem{Oemrawsingh2004}
S.S.R. Oemrawsingh, A. Aiello, E.R. Eliel, G. Nienhuis, and J.P. Woerdman, 
Phys. Rev. Lett., {\bf 92}, 217901, (2004).

\bibitem{Langford2004}
N.K. Langford, R.B. Dalton, M.D. Harvey, J.L. O'Brien, G.J. Pryde, A. Gilchrist, S.D. Bartlett, and A.G. White, 
Phys. Rev. Lett., {\bf 93}, 053601, (2004).

\bibitem{Lanyon2008b}
B.P. Lanyon, T.J. Weinhold, N.K. Langford, J.L. O'Brien, K.J. Resch, A. Gilchrist, and A.G. White, 
Phys. Rev. Lett., {\bf 100}, 060504, (2008).

\bibitem{O'Sullivan-Hale2005}
M. N. O'Sullivan-Hale, I. A. Khan, R. W. Boyd, and J. C. Howell, 
Phys. Rev. Lett., {\bf 94}, 220501, (2005).

\bibitem{Neves2005}
L. Neves, G. Lima, J.G. AguirreG\'{o}mez, C. H. Monken, C. Saavedra, and S. P\'{a}dua, 
Phys. Rev. Lett., {\bf 94}, 100501, (2005).

\bibitem{Lima2006}
G. Lima, L. Neves, I.F. Santos, J.G. AguirreG\'{o}mez, C. Saavedra, and S. P\'{a}dua, 
Phys. Rev. A, {\bf 73}, 032340, (2006).

\end{thebibliography}
% Produces the bibliography via BibTeX.

\end{document}